\documentclass[]{raa}            
\usepackage{graphicx,times}
\usepackage{natbib}

\providecommand{\bm}[1]{\mbox{\boldmath$#1$\unboldmath}}

\begin{document}

 \title{ Submillimeter/millimeter observations of the molecular clouds associated with the  Tycho' Supernova Remnant
}

 \volnopage{ {\bf 2010} Vol.\ {\bf } No. {\bf XX}, 000--000}
   \setcounter{page}{1}

   \author{Jin-Long Xu,
      \inst{1,2,3}
      \ Jun-Jie  Wang
      \inst{1,2}
      \ and Martin Miller
      \inst{4}}

    \institute{$^{1}$  National Astronomical Observatories, Chinese Academy of Sciences,
             Beijing 100012, China; {\it \small xujl@bao.ac.cn} \\
              $^{2}$  NAOC-TU Joint Center for Astrophysics, Lhasa 850000, China \\
              $^{3}$ Graduate University of the Chinese Academy of Sciences, Beijing,
             100080, China \\
             $^{4}$  I.Institute of Physics, University of Cologne, Cologne, 50937, Germany \\
             \vs \no
   {\small Received 2010 September 16; accepted 2010 December 10 }
}

 \abstract { We have carried out CO $J=2-1 $ and CO $J=3-2$ observations
toward Tycho's supernova remnant (SNR) using the  KOSMA
3m-Telescope. From these observations we identified three molecular
clouds (MCs) around the SNR. The small cloud in the southwest was
discovered for the first time. In the north and east, two MCs (cloud
A and cloud B) adjacent in space display a bow-shaped morphology,
and have broad emission lines, which provide some direct evidences
of the SNR-MCs interaction.  The MCs is revealed at -69$\sim$-59 km
s$^{-1}$, well coincident with the Tycho's SNR. The MCs associated
with Tycho's SNR have a mass of $\sim$ $2.13\times10^{3}\rm~
M_{\odot}$. Position-velocity diagrams show the two clouds adjacent
in velocity which means possible cloud-cloud collision in this
region. The maximum value (0.66 $\pm$ 0.10) of integrated CO line
intensity ratio ($I_{\rm CO}$$_ {J=3-2}$/$I_{\rm CO}$$_ {J=2-1}$)
for the three MCs agrees well with the previous measurement of
individual Galactic MCs, implying that the SNR shock just drove into
the MCs. The two MCs have a line intensity ratio gradient. The
distribution of the ratio appears to indicate that the shock
propagates from the southwest to the northeast.
   }

   \authorrunning{Xu et al. }            
   \titlerunning{Interaction between the Tycho' SNR and
 Molecular Clouds }  

   \maketitle
%
%
%
\section{INTRODUCTION}           
\label{sect:intro}

When a supernova  explodes near the molecular clouds (MCs), shock
generated by supernova remnant (SNR) can accelerate, compress, heat
 or even fragment surrounding interstellar MCs. They also can
enhance abundances with respect to quiescent cloud conditions of
different molecular species. So the SNR-MCs interaction plays a very
important role in the evolution of Interstellar Medium (Reynoso et
al. 2000). Moreover, the molecular lines observations of MCs
adjacent to SNRs can shed light on the SNRs' dynamical evolution and
physical properties. So far about half of the Galactic SNRs are
expected to be in physical associated with MCs (Reynoso $\&$ Mangum
2001). The association is indicated by morphological correspondence
of molecular emission, presence of molecular line broadening within
the extent of SNR, and presence of line emission with high
high-to-low excitation line intensity ratio, etc (see Jiang et al.
2010). However, only few cases about the SNR-MCs interaction has
been well confirmed. Even for some SNRs interacted with the
surrounding MCs, the detailed distribution of environmental
molecular gas is poorly known.

Tycho's SNR is known as a young type Ia SNR located in the Perseus
arm (Albinson et al. 1986), with an age of about 400 years. The
distance of Tycho's SNR is estimated to be 2.3 kpc (Kamper et al.
1978; Albinson et al. 1986; Strom 1988). The remnant has been widely
studied in various wavebands. From the neutral hydrogen HI
observation, Schwarz et al. (1995) concluded that the distance
should be 4.6$\pm$0.5 kpc; Reynoso et al. (1999) have performed an
HI absorption study toward the remnant, suggesting that Tycho's SNR
is currently interacting with a dense HI concentration in the
northeast ($V_{LSR}$=-51.5 km $s^{-1}$), which locally slows down
the expansion of the shock front. In radio continuum (Reynoso et al.
1997) and high-resolution X-ray wavebands  (Hwang et al. 2002), they
indicated that the radio continuum and X-ray emission along the
northeastern edge of  Tycho's SNR  is strongest. Based on the VLA
observation, Dickel et al. (1991) indicates that the northeastern
edge of Tycho's SNR is expanding with a very small and decreasing
velocity into the dense interstellar medium, and may interact with
it. Lee $\&$ Koo (2004) made the CO $J=1-0$ line observation towards
Tycho's SNR. They concluded that most of the CO $J=1-0$ emission
around Tycho's SNR is between -67 and -60 km $s^{-1}$, but the
velocity component in intervals -63.5  $\sim$ -61.5 km $s^{-1}$
appears to be in contact with the Tycho's SNR along its the
northeast boundary. Using CO $J=1-0$ line to study the environs of
Tycho's SNR, Cai et al. (2009) found that the CO $J=1-0$ emission
form a semi-closed molecular shell around the SNR, the emission in
velocity range of -69 $\sim$ -58 km $s^{-1}$ is associated with the
SNR. Because the calculated virial mass of clumps is greater than
their gravitational mass, they suggested that CO molecular clumps
are being violently disturbed by Tycho's SNR shock.

In order to understand the evolution of SNR interacting with MCs and
investigate the detailed distribution of the molecular gas around
Tycho's SNR, we have performed  CO $J=2-1 $ and CO $J=3-2$
observation toward Tycho' SNR. The observations cover for the first
time the whole area of Tycho's SNR in these frequencies. Due to the
observed molecular lines at higher frequencies, we can attain higher
angular resolution, which is critical  to identify relatively
compact core. Also, higher $J$ transitions are relatively more
sensitive to hot gas. Thus, we can detailedly understand the
distribution of the molecular gas in the vicinity of Tycho's SNR.

\section{Observations}

\label{sect:Obs} The observations of Tycho' SNR  were made in CO
$J=2-1$ and CO $J=3-2$ lines using the 3m  KOSMA sub-millimeter
telescope at Gornergrat, Switzerland in Jan, 2004. The half-power
beam widths of the telescope at observing frequencies of 230.538GHz
and 345.789GHz, are $130^{\prime\prime}$ and $80^{\prime\prime}$,
respectively. The pointing and tracking accuracy is better than
$10^{\prime\prime}$. The system temperature were about 120K. The
medium and variable resolution acousto optical spectrometers have
1501 and 1601 channels, with total bandwidth of 248 MHz and 544 MHz,
and equivalent velocity resolution of 0.21${\rm km\ s^{-1}}$ and
0.29${\rm km\ s^{-1}}$, respectively. The beam efficiency $B_{\rm
eff}$ is 0.68 and 0.72 at 230 GHz and 345 GHz, respectively. The
forward efficiency $F_{\rm eff}$ is 0.93. Line intensities were
converted on the  main beam
 scale, using the formula $T_{\rm mb}=F_{\rm
eff}/B_{\rm eff}\times T^\ast_{\rm A}$.

Mapping  observations are centered at RA(J2000)=$00^{\rm h}25^{\rm
m}22.7^{\rm s}$, DEC(J2000)=$64^{\circ}08'50.44^{\prime\prime}$
using the On-The-Fly mode, the total mapping area is
$14.5^{\prime}\times 15^{\prime}$ with a $0.5^{\prime}\times
0.5^{\prime}$ grid. The observed data were reduced using the CLASS
(Continuum and Line Analysis Single-Disk Software) and GREG
(Grenoble Graphic) software.

The 1.4GHz radio continuum emission data were obtained from the NRAO
VLA Sky Survey (NVSS; Condon et al. 1998).

\section{RESULTS}
\label{sect:data} Fig.1 (Left) shows  CO $J=2-1$ MCs distribution
around Tycho's SNR. The SNR appear as complete shell on the whole
and bright knots in southwest, which was detected by the 1.4 GHz
radio continuum emission. Three MCs were clearly identified around
Tycho's SNR. Each has been designated alphabetically, cloud A, cloud
B, and cloud C. In Fig.1 (Left), cloud A-C are well associated with
the 1.4 GHz radio continuum emission of Tycho's SNR. It appears that
the cloud A and cloud B are spatially adjacent. Here cloud C is for
the first time identified around Tycho. Fig.1 (Right) shows the CO
$J=2-1$ spectra towards the peak position of Cloud A-C ,
respectively. From these spectra, we can see that there are several
same velocity components in intervals -70 $\sim$ -53 and -7 $\sim$ 2
km $\rm s^{-1}$. The velocity component in interval -7 $\sim$ 2 km
$\rm s^{-1}$ should belong to the foreground source (Cai et al.
2009). A double-peaked feature is observed in intervals -70 $\sim$
-53 for cloud A and cloud B, which can be fitted by Gaussian,
respectively. Applying the rotation curve of Clemens (1985) together
with $V_{0}=220$ km $\rm s^{-1}$, where $V_{0}$ is the circular
rotation speed of the Galaxy, we obtained the distance of Cloud A-C.
Tycho's SNR exhibits a complete shell structure of about $\rm
8^{\prime}$. In view of the morphological correspondence of  Tycho's
SNR and the Cloud A-C, the tangent point (the center of the SNR) in
the direction to the SNR is  at $\sim$ 4.9 kpc, which is rather
close to the distance 4.6 $\pm$ 0.5 kpc obtained from the neutral HI
absorption observation (Schwarz et al. 1995). The fitted and derived
parameters for Cloud A-C are summarized in Table 1. In Table 1, the
clouds have broad emission lines. For cloud A, the distance of
component peaked at -56.8 km $\rm s^{-1}$ is 4.4 kpc, which is
smaller than that of Tycho's SNR ($\sim$ 4.9 kpc), so this velocity
component can be related to foreground Galactic gas emission. Other
components in intervals -70 $\sim$ -53 will be analyzed below in
relation with Tycho's SNR.

\begin{figure}
\includegraphics[width=80mm,angle=180]{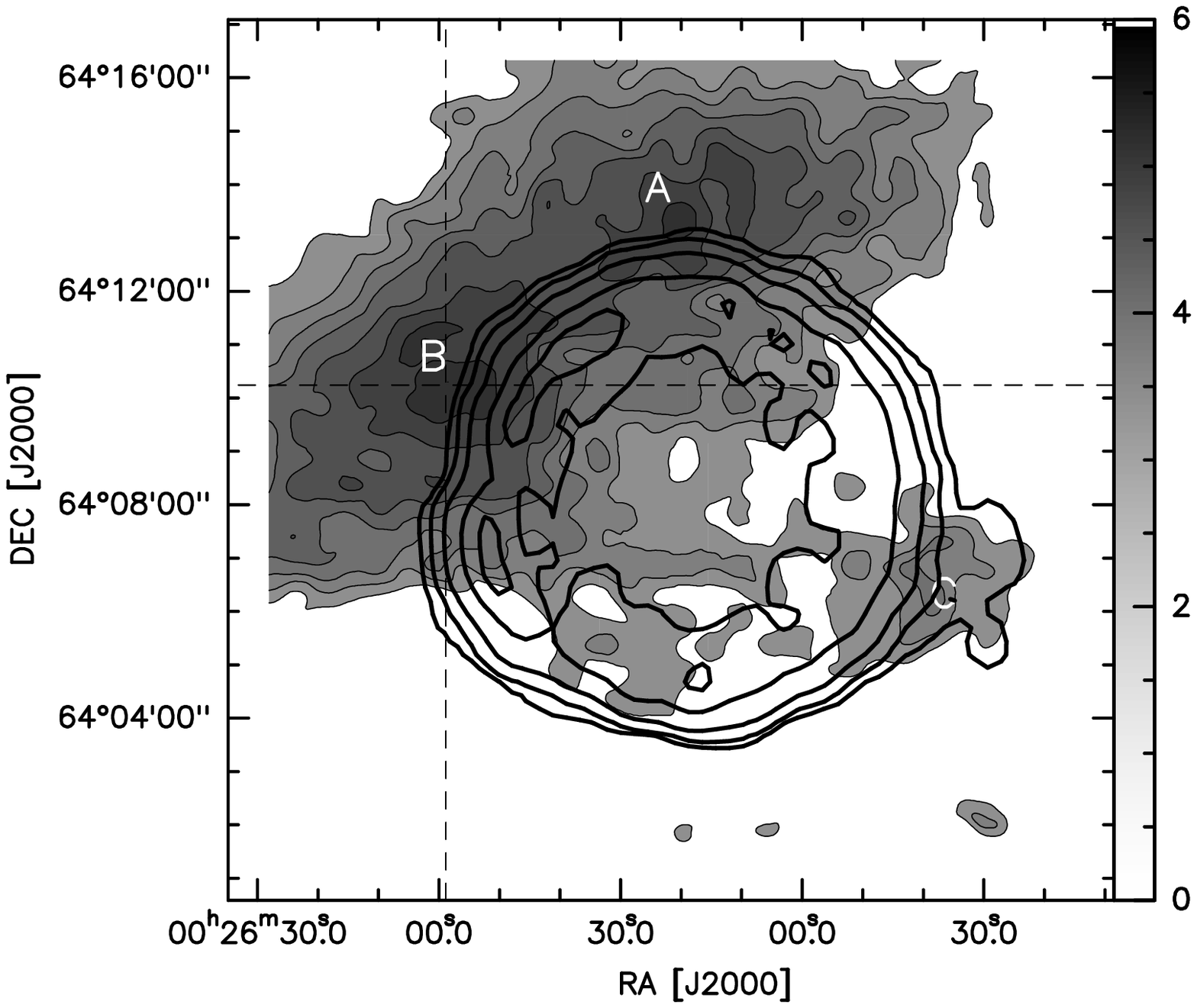}
\includegraphics[width=58mm,angle=-90]{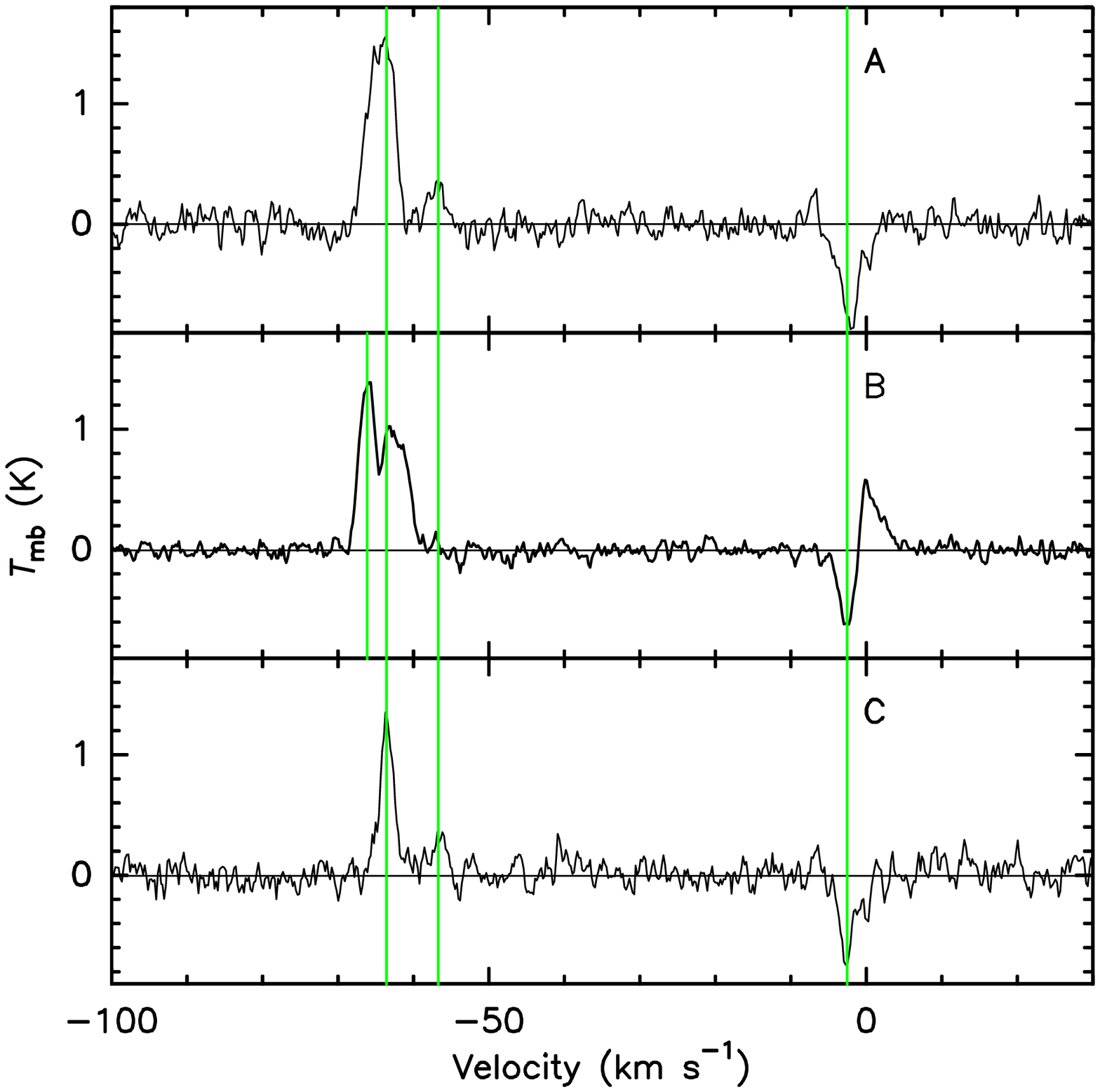}
\vspace{-60mm}\caption{Left: CO $J=2-1$ intensity map (gray scale)
integrated from -69 to -54 km s$^{-1}$, overlaid on the NVSS 1.4 GHz
radio continuum emission contours (black contours). Gray scale
levels are from 1.7 to 5.6 K km $\rm s^{-1}$ (the 1$\sigma$ value is
0.15 K km $\rm s^{-1}$ ). Two dashed lines mark the direction of the
cuts in Fig.4. Right: Spectra of molecular line CO $J=2-1$ at the
peak position of A-C cloud clumps. Letter A, B and C indicate the
different MC clumps. The vertical lines in the spectra mark the peak
velocity. } \label{Fig1}
\end{figure}

\begin{table}[!h]
\tabcolsep 2mm\caption[]{ The spectral line parameters at the peak
positions of clouds  }\vspace*{-14pt}
\def\temptablewidth{0.05\textwidth}%
  \begin{center}
  \begin{tabular}{cccccccc}
  \hline\noalign{\smallskip}
Name      & RA      & DEC                             & T$_{\rm mb}$ (rms)  & FW (rms)                & $V_{\rm LSR}$ (rms) & I (rms)                  & Distance\\
         & (h m s)&($\circ$ $\prime$ $\prime\prime$) &{\rm (K)} &${\rm (km\ s^{-1})}$ & ${\rm (km\ s^{-1})}$& (K ${\rm km\ s^{-1})}$&(kpc) \\
  \hline\noalign{\smallskip}
Cloud A   & 00 25 20.7 & 64 13 20.44   & 1.6(0.2)   & 9.7(0.2)    & -63.7(0.1) & 6.7(0.2) & 4.9\\  
          & 00 25 20.7 & 64 13 20.44   & 0.4(0.2)   & 6.6(0.5)    & -56.8(0.2) & 1.0(0.2) & 4.4\\  
Cloud B   & 00 25 38.7 & 64 10 20.44   & 1.1(0.1)   & 9.7(0.3)    & -63.5(0.1) & 4.5(0.2) & 4.9\\  
          & 00 25 38.7 & 64 10 20.44   & 1.2(0.1)   & 6.3(0.1)    & -66.2(0.1) & 3.1(0.3) & 5.0\\  
Cloud C   & 00 25 02.7 & 64 06 20.44   & 1.3(0.2)   & 7.0(0.1)    & -63.6(0.1) & 3.0(0.1) & 4.9\\  
\noalign{\smallskip}\hline
\end{tabular}\end{center}
\end{table}

\begin{figure}
   \centering
\vspace{-28mm}\includegraphics[width=15.0cm, angle=-90]{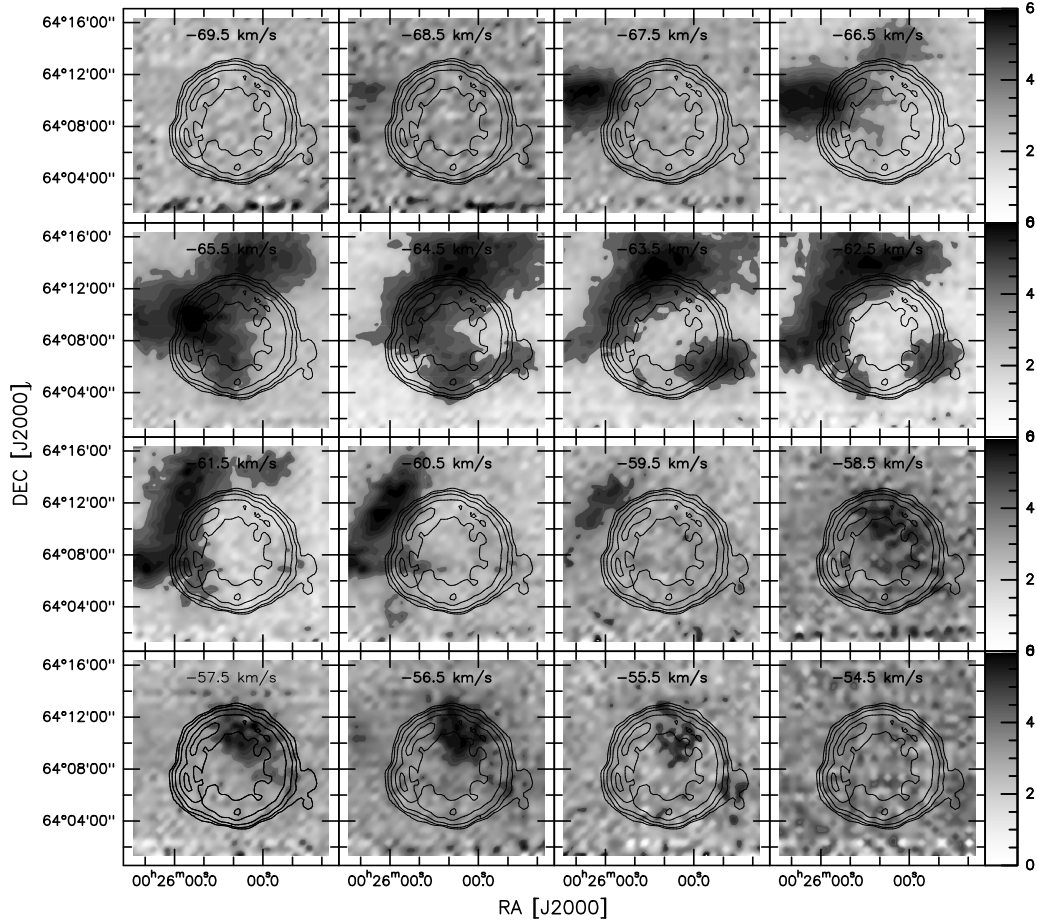}
   \caption{ CO $J=2-1$ channel maps each 1 km $s^{-1}$, overlapping the
NVSS 1.4 GHz radio continuum emission. The radio continuum emission
is indicated as black contours. Central velocities are indicated in
each image. }
   \label{Fig2}
   \end{figure}

The present high-$J$ transition observations enable us to map in
detail the molecular gas toward Tycho's SNR. After a careful
inspection of the CO component in intervals -70 $\sim$ -53, we find
that the only channel map can show the morphological and kinematical
possible signatures of interaction between Tycho's SNR and the
surrounding MCs. Fig.2 shows CO $J=2-1$ intensity channel maps over
the velocity range of -70 $\sim$ -53 km $\rm s^{-1}$ with interval 1
km $\rm s^{-1}$. The radio continuum emission is indicated as black
contours. There is virtually no CO $J=2-1$ emission at $V_{\rm LSR}$
$>$ -54 km $s^{-1}$ and $V_{\rm LSR}$ $<$ -69 km $\rm s^{-1}$, most
of the emission in between -69 km $\rm s^{-1}$ and -59 km $\rm
s^{-1}$. In addition, cloud A position has some faint emission in
intervals -59 $\sim$ -54 km $\rm s^{-1}$. The faint emission
corresponds  to a CO component peaked at -56.8 km $\rm s^{-1}$, as
seen from the CO $J=2-1$ spectra of cloud A in Fig.1(Right), which
can be related to foreground Galactic gas emission from above
analysis. Furthermore, cloud A position have some strong emission in
intervals -67 $\sim$ -59 km $\rm s^{-1}$, which can be related to a
CO component peaked at -63.7 km $\rm s^{-1}$. This component showing
an arc of molecular gas is striking coincident along the northeast
boundary. In Fig.2, there are two velocity components in the
intervals -69 $\sim$ -64 and  -67 $\sim$ -59 km $\rm s^{-1}$ for
cloud B, corresponding to peak at -66.2 km $\rm s^{-1}$ and -63.5 km
$\rm s^{-1}$ in spectra (see Fig.1), respectively. CO $J=2-1$
component at -69 $\sim$ -64 km $\rm s^{-1}$ appear to be in contact
with the remnant along its eastern boundary. In addition, we can
clearly see that the southwest CO $J=2-1$ molecular gas at -65
$\sim$ -61 km $\rm s^{-1}$ corresponds to cloud C.

The different transitions of CO trace different molecular
environment. In order to obtain the integrated intensity ratio of CO
$J=3-2$ and $J=2-1$ ($I_{\rm CO}$$_ {J=3-2}$/$I_{\rm CO}$$_
{J=2-1}$), we convolved the 80$^{\prime\prime}$ of CO $J=3-2$ data
with an effective beam size of
$\sqrt{130^{2}-80^{2}}=102^{\prime\prime}$ (Qin et al. 2008). The
integrated intensities were calculated for CO $J=2-1$ line in the
same velocity range as for CO $J=3-2$. The integrated range is from
-69 to -54 km $s^{-1}$. The rms noise level is 0.15 K km $\rm
s^{-1}$(1$\sigma$). Fig.3 shows the distribution of the ratio (color
scale) overlaid with the distribution of the CO $J=2-1$ line
integrated intensity (black dashed contours); The radio continuum
emission is indicated as black contours. The ratio values for the
cloud A, cloud B, and cloud C are between 0.23 and 0.66. The rms
level is  0.10 (1$\sigma$). The maximum value for the three MCs is
0.66, which is in agreement with the value measured in the Milky Way
(0.55 $\pm$ 0.08; Sanders et al. 1993), in NGC 253 (0.5 $\pm$ 0.1 in
the disk; Wall et al. 1991), and in M33 (0.69 $\pm$ 0.15; Wilson et
al. 1997). It also appear near value measured in the starburst
galaxies M82 (0.8 $\pm$ 0.2; G\"{u}sten et al. 1993). In Fig.3, the
high ratios (red color) are showed in the northeast part of cloud A
and B. The ratio value in this region is greater that in the region
where cloud overlaps with Tycho' SNR. The clouds have a ratio value
gradient increasing from southwest to northeast, suggesting that SNR
shock is expanding into cloud A and B, and start to compress the
clouds.

\begin{figure}
\centering\includegraphics[width=7cm, angle=-90]{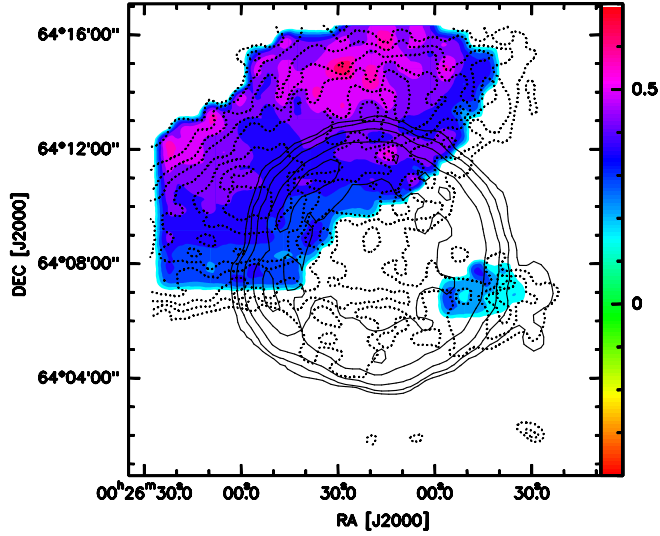} \caption{CO
$J=2-1$ intensity map (dashed line) are superimposed on the line
intensity ratio map (color scale), the line intensity ratios
($I_{\rm CO}$$_ {J=3-2}$/$I_{\rm CO}$$_ {J=2-1}$) range from 0.13 to
0.66 by 0.06. The rms level is  0.10 (1$\sigma$). The wedge
indicates the line intensity ratios scale. Black contours indicate
the NVSS 1.4 GHz radio continuum emission. }
   \label{Fig3}
   \end{figure}

Assuming local thermodynamical equilibrium (LTE) for the gas and
optically thick condition for CO $J=2-1$ line, we use the relation
$N_{\rm H_{2}}$ $\approx$ $10^{4}N_{\rm CO}$( Dickman 1978). The
column density is estimated as (Garden et al. 1991)
\begin{equation}
\mathit{N_{\rm CO}}= 6.9\times10^{14}\frac{T_{\rm ex}+0.92}{\rm
exp(-11.1/T_{\rm ex})}\int T_{\rm mb}dv ~ \rm cm^{-2},
\end{equation}

The excitation temperature ($T_{\rm ex}$) of the CO $J=2-1$ line is
estimated following the equation $T_{\rm ex}=11.1/{\ln[1+1/(T_{\rm
mb}/11.1+0.02)]}$, where $T_{\rm mb}$ is the corrected main beam
temperature. Furthermore, their mass is given by$M=\mu N_{\rm
H_{2}}S/(2.0\times10^{33})$, where the mean atomic weight of the gas
$\mu$ is 1.36, $S$ is the size of core region. The physical
parameters of the core are summarized in Table 2.

\begin{table}[]
  \tabcolsep 4mm\caption[]{ The physical parameters of the core in LTE. }\vspace*{-14pt}
  \label{Tab:publ-works}
  \def\temptablewidth{0.5\textwidth}
  \begin{center}\begin{tabular}{ccccccccc}
  \hline\noalign{\smallskip}
Name   &R (pc)& $T_{\rm ex}$(K) &$N_{\rm CO}$(cm$^{-2})$ & $N_{\rm H_{2}}$(cm$^{-2}$)& $M$($M_{\odot}$)               \\
  \hline\noalign{\smallskip}
Cloud A   &  5.1& 5.7 & 2.2$\times10^{17}$  & 2.2$\times10^{21}$ & 1151\\  
          &  1.4& 3.8 & 0.6$\times10^{17}$  & 0.6$\times10^{21}$ & 25\\  
Cloud B   &  2.8& 5.0 & 1.7$\times10^{17}$  & 1.7$\times10^{21}$ & 275\\  
          &  5.0& 5.1 & 1.1$\times10^{17}$  & 1.1$\times10^{21}$ & 581\\  
Cloud C   &  2.1& 5.3 & 1.1$\times10^{17}$  & 1.1$\times10^{21}$ & 96\\  
\noalign{\smallskip}\hline
\end{tabular}\end{center}
\end{table}

\section{DISCUSSION}
\label{sect:discussion} Tycho's SNR has complete radio shell on the
whole and bright radio knots in southwest. Among the MCs surrounding
Tycho's SNR, we have seen that three clouds (cloud A, cloud B and
cloud C) are spatially coincident with the SNR. Broad emission lines
detected in cloud A and B and bow-shaped morphology suggest that the
MCs may be interacting with the SNR. After a careful inspection of
the CO component, we find that CO component of cloud A in intervals
-67 $\sim$ -59 km $\rm s^{-1}$ (peaked at -63.7 km $\rm s^{-1}$) and
CO component of cloud B in intervals -69 $\sim$ -64 km $\rm s^{-1}$
(-66.2 km $\rm s^{-1}$) and -67 $\sim$ -59 km $\rm s^{-1}$ (-63.5 km
$\rm s^{-1}$) are well associated with Tycho's SNR. The
position-velocity diagram across the peak of cloud B along the
north-south and the east-west direction, respectively, are
constructed from the CO $J=2-1$ line (see Fig.4). In Figure 4,
obviously, the velocity components of cloud A and cloud B are
adjacent. For cloud A, CO component in intervals -67 $\sim$ -59 km
$\rm s^{-1}$ may belong to cloud B, because cloud A and cloud B are
adjacent in space and in velocity, and have the approximately same
peak velocity in spectra. We concluded that cloud A and cloud B may
not only simply be overlapping along the line of the sight, but also
colliding each other. Furthermore, the new cloud C is also
associated well with bright knots, the bright knots may be radio
emissivity increase due to compression of the shocked material,
suggesting that cloud C appears to be swept up by the SNR shock in
the southwest area.

\begin{figure}
\includegraphics[width=48mm,angle=-90]{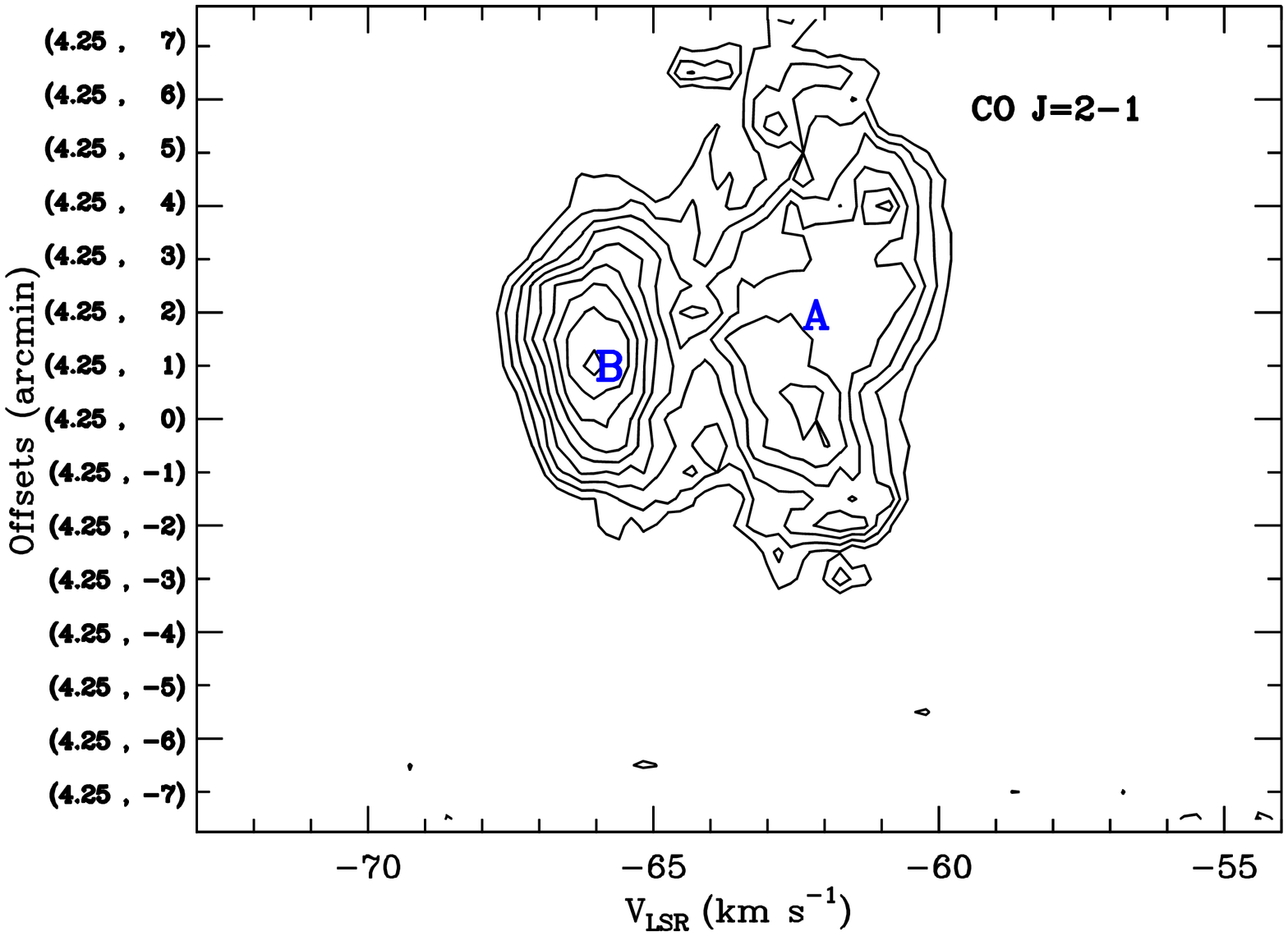}
\includegraphics[width=48mm,angle=-90]{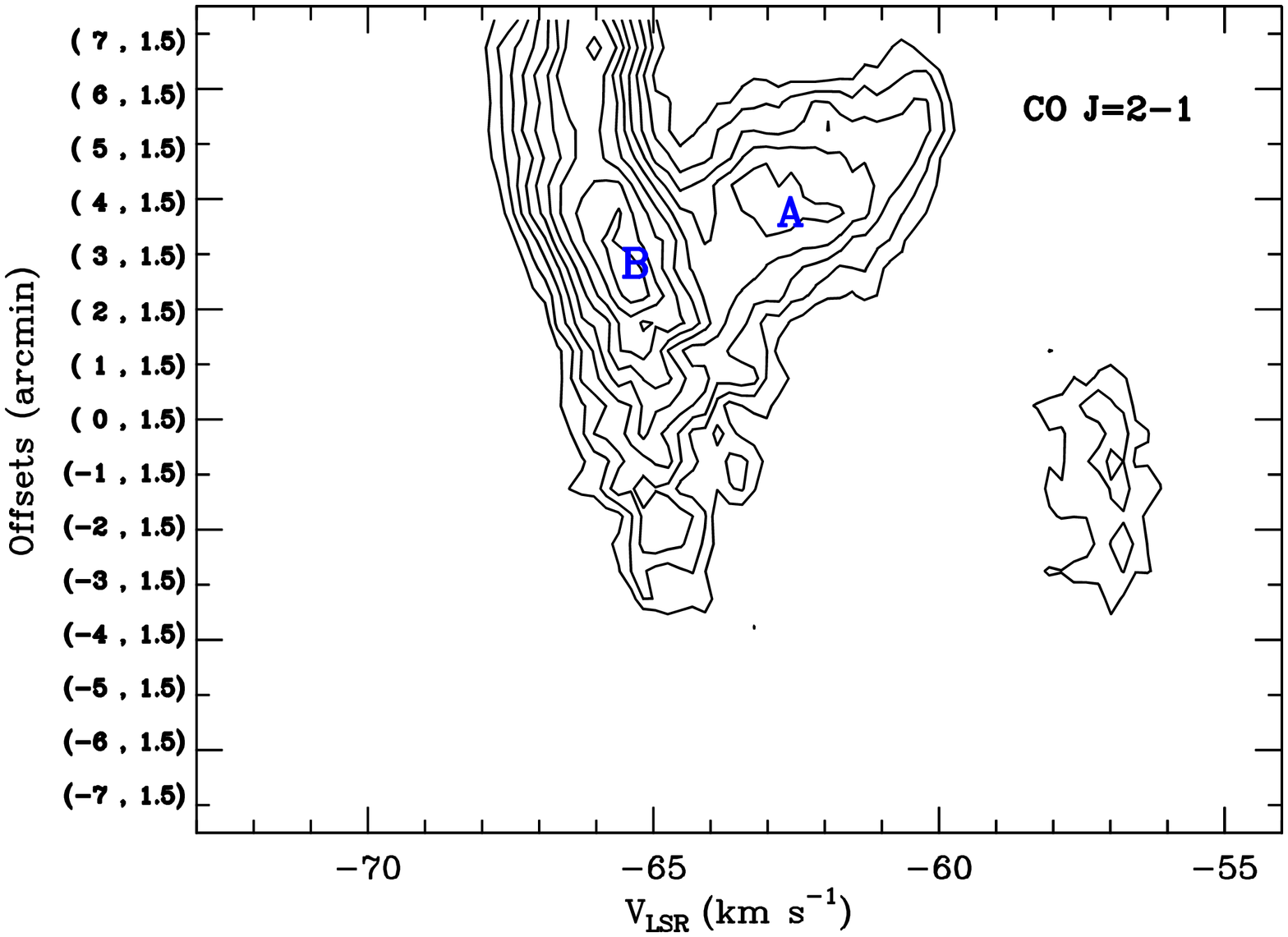}
\vspace{0mm}\caption{\small  P-V diagram constructed from the CO
$J=2-1$ transition for Cloud B. Left panel: Contour levels are from
0.3 K to 4.4 K by the 0.1 K, with a cut along the north-south
direction. Right panel: Contour levels  are the same as in left
panel, with a cut along the east-west direction. } \label{Fig3}
\end{figure}

The maximum value of integrated CO line intensity ratio for the
three MCs is 0.66 $\pm$ 0.10. The comparison with previously
published observations reveals that the $I_{\rm CO}$$_
{J=3-2}$/$I_{\rm CO}$$_ {J=2-1}$ for the MCs associated with Tycho's
SNR  agrees well with the value measured in in the Milky Way (0.55
$\pm$ 0.08; Sanders et al. 1993), in NGC 253 (0.5 $\pm$ 0.1 in the
disk; Wall et al. 1991), and in M33 (0.69 $\pm$ 0.15; Wilson et al.
1997). It also appear near value measured in the starburst galaxies
M82 (0.8 $\pm$ 0.2; G\"{u}sten et al. 1993). The high $I_{\rm CO}$$_
{J=3-2}$/$I_{\rm CO}$$_ {J=2-1}$ in starburst galaxies may be due to
unusual conditions in these dense and hot regions (Aalto et al.
1997), while for normal MCs the most likely explanation is a
significant contribution to the CO emission by low column density
material (Wilson $\&$ Walker 1994). Moreover, the high $I_{\rm
CO}$$_ {J=3-2}$/$I_{\rm CO}$$_ {J=2-1}$ value (3.4) in the MCs
interacting with SNR IC443 (Xu et al. 2010) exceed previous
measurement of individual Galactic MCs, implying that the SNR shock
has driven into the MCs. For the MCs around Tycho's SNR, the $I_{\rm
CO}$$_ {J=3-2}$/$I_{\rm CO}$$_ {J=2-1}$ value (0.66 $\pm$ 0.10 ) may
indicate that the Tycho's SNR shock just drove into the surrounding
MCs. In addition, the MCs associated with HII regions have higher
ratio of $I_{\rm CO}$$_ {J=3-2}$/$I_{\rm CO}$$_ {J=2-1}$ and have
higher gas temperature than those sources without HII regions,
indicating that the high line ratio may be due to heating of the gas
by the massive stars (Wilson et al. 1997). Different masers maybe
occur in different astrophysical environment. The $\rm H_{2}O$
masers are located near the MSX sources and within the maximum line
intensity ratio $I_{\rm CO}$$_ {J=3-2}$/$I_{\rm CO}$$_ {J=2-1}$
regions, suggesting that $\rm H_{2}O$ masers occur in relatively
warm environments (Qin et al. 2008). Thus, we concluded that the
line intensity ratios based on the optically thick CO transitions
indicate the temperature varies at different positions. The MCs
associated with Tycho' SNR have a  ratio value gradient increasing
from southwest to northeast, may imply that a shock has driven into
MCs and  compressed MCs. Since the MCs are compressed, it lead to
the temperature of molecular gas increases. Hence the line intensity
ratio $I_{\rm CO}$$_ {J=3-2}$/$I_{\rm CO}$$_ {J=2-1}$ will increase.
We consider that high $I_{\rm CO}$$_ {J=3-2}$/$I_{\rm CO}$$_
{J=2-1}$ is also identified as one good signature of SNR-MCs
interaction system. From Table 1, the total mass of MCs is $\sim
2.13\times10^{3}$.

\section{SUMMARY}
\label{sect:conclusion} We have presented large-area map of
molecular clouds (MCs) in the vicinity of Tycho' SNR  in CO $J=2-1$
and CO $J=3-2$ lines.  The complete map suggests that MCs are mainly
distributed along the northern and eastern  boundary of the SNR. It
also reveals a new cloud (cloud C) that appears to be swept by the
SNR shock in the southwestern area. The bow-sharped morphology, the
broad emission lines, as well as the integrated CO line intensity
ratios ($I_{\rm CO}$$_ {J=3-2}$/$I_{\rm CO}$$_ {J=2-1}$) (0.66 $\pm$
0.10) further suggest that these MCs are interacting with Tycho'
SNR. The MCs is revealed at -69$\sim$-59 km s$^{-1}$, well
coincident with the Tycho's SNR. The MCs associated with Tycho's SNR
have a mass of $\sim$ $2.13\times10^{3}\rm~ M_{\odot}$. In the
northern and eastern MCs, we find a line intensity ratio gradient
along the southwest-northeast direction, implying that Tycho' SNR
shock has driven into the MCs. We concluded that the line intensity
ratios based on the optically thick CO transitions indicate the
temperature varies at different positions. Both the integrated
intensity maps and position-velocity diagrams show the eastern and
northern clouds adjacent in space and velocity. Together with the
approximately same peak velocity in spectra, we found that the two
MCs are interacting, indicating possible cloud-cloud collision in
this region.

\begin{acknowledgements} We are very grateful to the anonymous referee for his/her helpful comments and suggestions.
 This work was supported by the National Natural Science Foundation of China under Grant No.10473014.
\end{acknowledgements}


\end{document}